\documentclass[ aps, prx, twocolumn, showpacs,nofootinbib,superscriptaddress,longbibliography, 10pt]{revtex4-2}
\usepackage{lipsum}% http://ctan.org/pkg/lipsum
\usepackage{graphicx} % Required for inserting images
\usepackage{amsmath}
\usepackage{amssymb}
\usepackage{bm}
\usepackage{mathtools}
\usepackage{MnSymbol}
\usepackage{amsthm}
\usepackage{braket}
\usepackage{comment}
\usepackage{xcolor}
\usepackage{acronym}
\usepackage{float}
\makeatletter
\let\newfloat\newfloat@ltx
\makeatother
\usepackage{algorithm}
\usepackage{algpseudocode}
\usepackage{hyperref}

\graphicspath{ {./Figures}}
\usepackage[protrusion=true,expansion=true]{microtype}
\usepackage{xurl}

\begin{document}

\title{Active Learning for Calibrating Entangling Gates via Surrogate-Based Optimization}

\author{Caleb Walton}
\affiliation{Department of Electrical and Computer Engineering, University of Washington, Seattle, WA 98195, USA}

\author{Patricia García-Caspue\~{n}as}
\affiliation{Department of Mechanical Engineering, University of Washington, Seattle, WA 98195, USA}
\affiliation{NSF AI Institute in Dynamic Systems, University of Washington, Seattle, WA 98195, USA}

\author{Filippo Zacchei}
\affiliation{MOX - Department of Mathematics, Politecnico di Milano, Piazza Leonardo da Vinci 32, 20133 Milano, Italy}
\affiliation{Pasteur Labs, Brooklyn, NY, USA}

\author{Ana Larra\~{n}aga}
\affiliation{Department of Mechanical Engineering, University of Washington, Seattle, WA 98195, USA}
\affiliation{NSF AI Institute in Dynamic Systems, University of Washington, Seattle, WA 98195, USA}

\author{Melissa C. Revelle}
\affiliation{Sandia National Laboratories, Albuquerque, New Mexico, 87185, USA}

\author{Edward C. Tortorici}
\affiliation{Sandia National Laboratories, Albuquerque, New Mexico, 87185, USA}

\author{Brian K. McFarland}
\affiliation{Sandia National Laboratories, Albuquerque, New Mexico, 87185, USA}

\author{Daniel S. Lobser}
\affiliation{Sandia National Laboratories, Albuquerque, New Mexico, 87185, USA}

\author{Christopher G. Yale}
\affiliation{Sandia National Laboratories, Albuquerque, New Mexico, 87185, USA}

\author{Susan M. Clark}
\affiliation{Sandia National Laboratories, Albuquerque, New Mexico, 87185, USA}

\author{Steven L. Brunton}
\affiliation{Department of Mechanical Engineering, University of Washington, Seattle, WA 98195, USA}
\affiliation{NSF AI Institute in Dynamic Systems, University of Washington, Seattle, WA 98195, USA}

\author{Sara Mouradian}
\affiliation{Department of Electrical and Computer Engineering, University of Washington, Seattle, WA 98195, USA}

\begin{abstract}

High-fidelity quantum gates require precise tuning of many classical control parameters, and drift in these parameters necessitates periodic recalibration that takes time and qubit resources. Physics-based calibration can be efficient, but it requires a validated and high-fidelity model of the quantum gate under the given classical control and environment. A high-fidelity model is expensive to build, especially as systems scale, and will have to be periodically re-validated as system parameters drift. Here, we present a physics-agnostic active-learning framework that identifies optimal control parameters from noisy measurement outcomes. A generic Gaussian process with a heteroskedastic noise model set by quantum projection noise is used to model the score landscape of the calibration sequence. Bayesian optimization simultaneously learns the calibration landscape and locates its optimum. We validate the technique by calibrating a two-qubit trapped-ion entangling gate. In simulation, where the ground-truth landscape is known, we find that the surrogate accurately models the three-parameter landscape and the calibrated infidelity saturates at a level set by the number of shots per evaluation. Validation on the Sandia Quantum User Testbed shows that the physics-agnostic protocol works on an experimental system with four calibrated parameters, needing an average of 44$\pm 11$ evaluation points at 100 shots per point to model the calibration landscape and find the optimal parameter set. These results show that active learning with physics-agnostic surrogate models can calibrate quantum gates.

\end{abstract}
\acresetall
\maketitle

\acrodef{BO}[BO]{Bayesian Optimization}
\acrodef{GP}[GP]{Gaussian Process}
\acrodefplural{GP}[GPs]{Gaussian Processes}
\acrodef{GPR}[GPR]{Gaussian Process Regression}
\acrodef{MS}[MS]{M\o lmer-S\o rensen}
\acrodef{UCB}[UCB]{Upper Confidence Bound}
\acrodef{EI}[EI]{Expected Improvement}
\acrodef{RMSE}[RMSE]{Root Mean Squared Error}
\acrodef{MSSE}[MSSE]{Mean Squared Scaled Error}

\section{Introduction}\label{sec:intro}

High-fidelity quantum gate operations are a prerequisite for near-term quantum advantage~\cite{Preskill2018NISQ} and fault-tolerant quantum error correction~\cite{Fowler2012SurfaceCodes}. Regardless of the platform, achieving high-fidelity gates requires the precise calibration of multiple classical control parameters that define the implemented Hamiltonian. Environmental drift necessitates periodic re-calibration~\cite{Proctor2020Drift}, which consumes time and qubit resources that could otherwise be used for computation. This overhead grows with the number of gates that must be maintained. Thus, fast and hands-free calibration routines will be a necessary component in any quantum technology.

Given a faithful model of the quantum system and its classical control infrastructure, Bayesian calibration has been used to rapidly tune entangling operations~\cite{innsbruckBayesian}. However, this approach requires classical simulations of the gate dynamics, which become computationally expensive as the number of control parameters grows. Moreover, this depends on a sufficiently precise theoretical model of the device. Though this can, in principle, be estimated through measurements of the quantum system~\cite{Nielsen2021GST,Yu2023HamiltonianLearning, Mohseni2008QPT, OzeriQPT}, these methods scale unfavorably with system size and complexity, especially as the characterization must be done over a range of input parameters and the model may drift, requiring periodic re-validation.  Thus, as quantum processors scale to more qubits and deeper circuits, calibration routines must remain efficient and operate even without an exact model of the experimental system.

Sequential sweeps of physical control parameters are intuitive and need no prior knowledge, but the number of measurements needed grows exponentially with the number of parameters. More sophisticated techniques have been proposed~\cite{superconductingNelderMead, Qian2026Aug, Mazurek2025Jul, Wittler2021Mar, Egger2014Jun, Jeanette2025Jan, Maksymov2021Jun, Magann2025Dec, Rausell-Campo2026Jul} for fast calibration models of varying degrees of imperfection. A robust and scalable calibration procedure should be global and robust to noise, make full use of every measurement, and need as little prior information as possible.

Here, we draw inspiration from classical engineering domains, where data-driven surrogates enable control and optimization of complex systems~\cite{brunton2025machine}. Surrogate-based \ac{BO} is a natural framework for quantum systems as it can accommodate unavailable gradients, and noisy and costly evaluations~\cite{booker1998design,Jones1998,Frazier2018}. \ac{BO} selects new evaluation points by balancing exploration and exploitation in a sample-efficient manner~\cite{Snoek2012}. Compared to exhaustive sweeps and local search routines, \ac{BO} can use previous evaluations more effectively to identify promising regions of the parameter space with fewer queries, while accounting for parameter-dependent uncertainty in the measured objective~\cite{Snoek2012,Frazier2018}. 

Different surrogate models have different strengths and weaknesses. Parametric Bayesian models such as the Bayesian extensions~\cite{fung2025,hirsh2022sparsifying,fasel2022ensemble} of SINDy~\cite{brunton2016} or Polynomial Chaos Expansion~\cite{berkemeier2023accelerating}, provide robust uncertainty quantification but require choosing a library of basis functions in advance based on a priori knowledge of the system. Neural network surrogates~\cite{pang2019neural} are also poorly matched to calibration of quantum systems where measurements are inherently probabilistic and expensive. In contrast, \acp{GP}~\cite{williams2006gaussian} are non-parametric, give exact analytical posterior predictions, and are flexible in adapting to new training data despite observation noise. They are the standard surrogate for \ac{BO}~\cite{shahriari2015taking}, with demonstrated success across a variety of experimental settings~\cite{morita2022applying,chen2024machine,hase2018phoenics, burger2020mobile,fan2019robotic}.

\begin{figure*}[t]
    \centering
    \includegraphics[width=\linewidth]{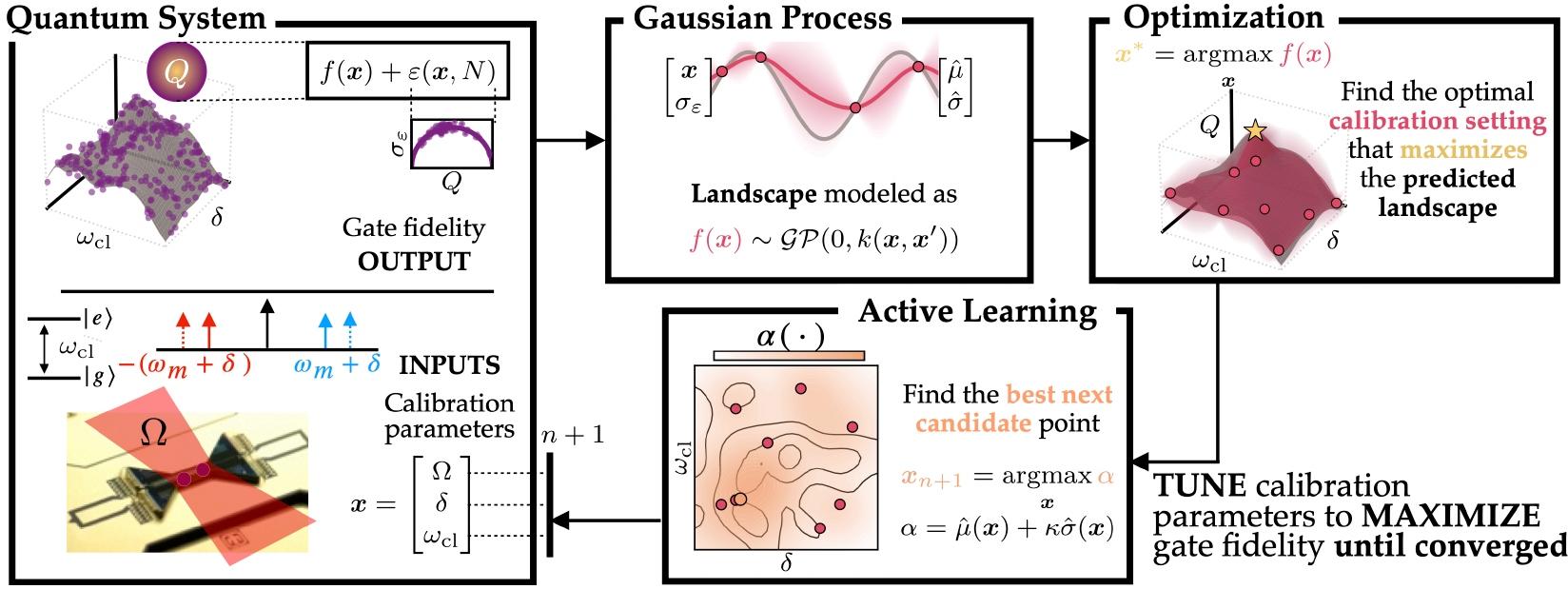}
    \caption{\textbf{Schematic of the optimization pipeline}. The goal is to maximize the gate fidelity score $Q$ as a function of the calibration parameters: Rabi frequency $\Omega$,  sideband detuning $\delta$ and center-line detuning $\omega_{\mathrm{cl}}$. First, the score is queried from the quantum system, represented by an unknown noise-free landscape $f(\bm{x})$ and a  modeled noise term $\varepsilon$ dependent on the inputs and the number of measurement shots $N$. Next, a Gaussian Process (GP) acts as a probabilistic surrogate that models $f(\bm{x})$ from the available system scores to predict the mean and variance of unexplored calibration settings. This surrogate is exploited for fast optimization of the predicted mean landscape. Finally, an active learning step selects the best next calibration setting and the process is repeated until convergence.}
    \label{fig:drawing}
\end{figure*}

In this work, we construct a probabilistic surrogate model based on a \ac{GP} that is agnostic to the underlying physics of the system. As pictured in Fig.~\ref{fig:drawing}, we simultaneously learn the landscape of the quantum system with respect to the classical control parameters and identify its high-fidelity regions through noisy measurements of the quantum system. As measurement noise may vary across the control landscape, we use a heteroskedastic noise model, where different control settings present different measurement uncertainties~\cite{makarova2021}.

We validate the approach in simulation and experiment by calibrating multiple physical controls of a trapped-ion \ac{MS} gate with an iterative workflow described schematically in Fig.~\ref{fig:drawing}, demonstrating that a \ac{GP} can capture the parameter-dependent landscape. Active learning then allows us to efficiently identify the optimal parameters, even in the presence of quantum projection noise. 

In Sec.~\ref{sec:methods:BO} we describe the \ac{GP} surrogate and the active learning \ac{BO} loop. Section~\ref{sec:methods:quantum} describes the simulated Hamiltonian, the parameters of interest and noise model.  Section~\ref{sec:results:GP} shows the performance of the surrogate, analyzing how model accuracy scales with the number of training points and the measurement shots taken per point. In Sec.~\ref{sec:results:opt_1} and \ref{sec:results:opt_2} we add active learning and evaluate how the optimizer converges under different parameter ranges and finite-shot budgets, identifying the fidelity saturation limits set by quantum projection noise. Finally, Sec.~\ref{sec:results:exp} evaluates the performance of the technique on QSCOUT, validating the experimental performance.

\section{Surrogate modeling for Bayesian Optimization} \label{sec:methods:BO}

We introduce the mathematical framework responsible for accelerating our calibration protocol. A \ac{GP} is used as a fast approximation of the system physics over the control parameter space. In our proposed optimization process, we sequentially update the \ac{GP} surrogate with a new measurement. For each iteration, \ac{BO} selects this new sample as the most informative control set to jointly improve our understanding of the system and rapidly drive convergence toward the desired tuning goal.

\subsection{Gaussian Process Regression}

The surrogate acts as a computationally low-cost proxy for the noisy quantum experiment. Given a set of previously tested calibration parameters and their measured outcomes, the surrogate can predict the expected measurement at new parameter values while simultaneously quantifying the uncertainty of that prediction. The approach relies on a \ac{GP} \cite{williams2006gaussian,murphy2012machine}, which defines a probability distribution over possible functions consistent with the observed data. This makes it flexible enough to model an unknown calibration landscape without assuming a fixed parametric form.

The measured physical observable (e.g., quantity of interest) is modeled as a noisy evaluation $Q_i$ of an
unknown calibration landscape $f(\bm{x})$,
\begin{equation}
    Q_i = f(\bm{x}_i) + \varepsilon_i,
    \qquad
    \varepsilon_i
    \sim
    \mathcal{N}
    \left(
    0,
    c^2\sigma_{\varepsilon}^2(\bm{x}_i, N)
    \right),
    \label{eq:GP_noise}
\end{equation}
where $\bm{x}_i\in\mathbb{R}^d$ is the $d$-dimensional vector of experimental calibration parameters. The additive noise term $\varepsilon_i$ changes for each experimental setting $\bm{x}_i$ and is assumed to follow a zero-mean Gaussian distribution, where its variance $\sigma_{\varepsilon}^2(\bm{x}_i, N)$ is prescribed by the quantum projection noise~\cite{QPN} after $N$ measurements. Therefore, the observation noise is heteroskedastic. The global scaling parameter $c$ is the overall magnitude of this noise and is one of the hyperparameters optimized during the surrogate fitting process.

In \ac{GPR}, the first step is to place a prior distribution on the unknown function $f$, namely, 
\begin{equation}
    f(\bm{x})\sim \mathcal{GP}\left(0, k(\bm{x}, \bm{x}')\right).
    \label{eq:GP_model}
\end{equation}
In this way, the unknown physical landscape is assumed to be a normally distributed random variable with zero mean and a covariance structure specified through a kernel function $k(\bm{x},\bm{x}')=\operatorname{cov}(f(\bm{x}),f(\bm{x}'))$. This covariance between outcomes at different settings encodes how the calibration landscape is expected to vary across the $d$-dimensional input space.

Because both the prior and the independent measurement noise model are Gaussian, conditioning the model on observed experimental data gives a Gaussian posterior. That is, predictions at any new control setting $\bm{x}$ are analytically resolved via a closed-form predictive distribution, characterized by a posterior mean and variance.

The training set after $m$ experimental evaluations is
\begin{equation}
    \mathcal{D}
    =
    \left\{
    \left(
    \bm{x}_i,
    Q_i,
    \sigma_{\varepsilon,i}^2
    \right)
    \right\}_{i=1}^{m},
    \qquad
    \sigma_{\varepsilon,i}^2
    =
    \sigma_{\varepsilon}^2(\bm{x}_i, N).
\end{equation}
We collect the inputs in the matrix $\bm{X}\in\mathbb{R}^{m\times d}$, the measured scores in the vector $\bm{Q}\in\mathbb{R}^m$, and the noise variances in the diagonal matrix,
\begin{equation}
    \bm{\Lambda}
    =
    \operatorname{diag}
    \left(
    \sigma_{\varepsilon,1}^2,
    \ldots,
    \sigma_{\varepsilon,m}^2
    \right).
\end{equation}
The diagonal matrix $\bm{\Lambda}$ accounts for the varying measurement uncertainties associated with different experimental configurations. The presented heteroskedastic \ac{GP} model differs from fully unconstrained heteroskedastic GP models, where the noise landscape is inferred from data without an a priori structure \cite{le2005heteroscedastic}, and is related to stochastic  models with input-dependent measurement variance \cite{ankenman2010stochastic}.

The predictive distribution for a new, unexplored configuration $\bm{x}$ is given by conditioning the \ac{GP} prior on the collected measurements,
\begin{equation}
    f(\bm{x})
    \mid
    \bm{Q},\bm{X},\bm{x}
    \sim
    \mathcal{N}
    \left(
    \hat{\mu}(\bm{x}),
    \hat{\sigma}^2(\bm{x})
    \right), \label{eq:posterior}
\end{equation}
where
\begin{equation}
    \begin{cases}
    \hat{\mu}(\bm{x})
    =
    \hat{\bm{k}}^\top
    \left(
    \bm{K}
    +
    c^2\bm{\Lambda}
    \right)^{-1}
    \bm{Q},
    \\[0.2cm]
    \hat{\sigma}^2(\bm{x})
    =
    k(\bm{x}, \bm{x})
    -
    \hat{\bm{k}}^\top
    \left(
    \bm{K}
    +
    c^2\bm{\Lambda}
    \right)^{-1}
    \hat{\bm{k}}. \label{eq:posterior_mean_s}
    \end{cases}
\end{equation}
Here, $\hat{k}_j = k(\bm{x}_j, \bm{x})$, and $K_{i,j} = k(\bm{x}_i, \bm{x}_j)$. Note that $\hat{\bm{k}}\in\mathbb{R}^{m}$ is the covariance vector between the measured configurations and the new target, while $\bm{K}\in\mathbb{R}^{m\times m}$ is the covariance matrix constructed from the previously measured control configurations. In Eq.~\ref{eq:posterior_mean_s}, the varying noise information encoded in $\bm{\Lambda}$ affects the covariance matrix of the noisy observations $\bm{K}+c^2\bm{\Lambda}$, and consequently, the mean and variance of the prediction.

Kernel selection dictates the assumed smoothness and correlation structure of the modeled physical landscape, governing the model's interpolation capabilities. We use the Matérn 3/2 kernel with Automatic Relevance Determination, which is particularly effective for physical systems~\cite{fan2019robotic}. By assigning a characteristic length scale $\ell_j$ to each input dimension, this approach provides access to more expressive correlation structures. It flexibly adapts to both rough and smooth landscapes, while granting control over local and global structure. The kernel is
\begin{equation}
    k(\bm{x}, \bm{x}') = \sigma_f^2 \left( 1 + \sqrt{3} r(\bm{x}, \bm{x}') \right) \exp \left( -\sqrt{3} r(\bm{x}, \bm{x}') \right).
\end{equation}
Here, $\sigma_f^2$ represents the signal variance, and the scalar distance between two input configurations is defined as 
\begin{equation}
r(\bm{x}, \bm{x}') = \sqrt{ \sum_{j=1}^d \left( \frac{x_j - x_j'}{\ell_j} \right)^2 }. 
\end{equation}
Consequently, the complete set of hyperparameters governing the kernel and the structured noise model has dimensionality $d+2$, denoted as $\bm{\Theta}=[c, \sigma_f, \ell_1, \dots, \ell_d]^\top $.
The hyperparameters are estimated by minimizing the negative log-marginal likelihood, also known as type-II maximum likelihood or empirical Bayes \cite{williams2006gaussian}:
\begin{equation}
    \bm{\Theta}^* = \underset{\bm{\Theta}}{\operatorname{arg min}} [-\log{P(\bm{Q}|\bm{X}, \bm{\Theta})}], \label{eq:MLE1}
\end{equation}
where
\begin{equation}
\begin{split}
    -\log{P(\bm{Q}|\bm{X}, \bm{\Theta})} &= \frac{1}{2} \bm{Q}^\top \left(\bm{K} + c^2 \bm{\Lambda}\right)^{-1} \bm{Q} + \\
    &\quad + \frac{1}{2} \log |\bm{K} + c^2 \bm{\Lambda}| + \frac{m}{2} \log(2\pi).
\end{split}
\label{eq:MLE}
\end{equation}
This criterion balances data fit and model complexity. In Eq.~\ref{eq:MLE}, the first term governs how well the model aligns with the measured data, while the second two terms act as a complexity penalty and normalization such that fluctuations due to noise are not interpreted as true physical features. The resulting objective is differentiable with respect to the hyperparameters and can therefore be optimized efficiently using gradient-based methods, typically with multiple initializations to mitigate local minima.
%############################################################################
%\vspace{-0.4cm}
\subsection{Bayesian Optimization}
%\vspace{-0.4cm}
%############################################################################
In GPR, the predicted fidelity at any given experimental configuration is modeled as a normal distribution. The analytic tractability of both the mean and the predictive variance is a cornerstone of \ac{BO} \cite{Jones1998,shahriari2015taking}. The objective is thus to find the optimal calibration setting $\bm{x}^*$ that maximizes the true, unobservable physical landscape $f(\bm{x})$,
\begin{equation}
\bm{x}^* = \underset{\bm{x}}{\operatorname{arg max}} \: f(\bm{x}).
\end{equation}
\ac{BO} serves as a gradient-free global optimization method for experimental systems where measurements are expensive and noisy. It uses the probabilistic approximation provided by the surrogate model to sequentially propose new configurations for querying the physical process. The primary goal is to locate the global optimum in the fewest possible physical evaluations by intelligently balancing the exploration of highly uncertain regions with the exploitation of promising regions.

\begin{algorithm}[t]
\caption{Active learning framework for measurement-efficient heteroskedastic Bayesian optimization.}
\label{alg:hbo}
\begin{algorithmic}[1]
\Require Initial training dataset $\mathcal{D}_0 = \{(\bm{x}_i, Q_i, \sigma^2_{\varepsilon,i})\}_{i=1}^m$, evaluation budget $n_{max}$, acquisition function $\alpha(\cdot)$, hyperparameter update interval $h$.
\State $n \leftarrow 1$
\While{$n \leq n_{max}$}
    \State \textbf{1. Hyperparameter optimization:} 
    \If{$(n-1) \pmod{h} = 0$}
        \State $\bm{\Theta}_{n-1} \leftarrow \underset{\bm{\Theta}}{\operatorname{arg min}} \left[ -\log P(\bm{Q}_{n-1} \mid \bm{X}_{n-1}, \bm{\Theta}) \right]$
        \State $f(\bm{x}) \mid \mathcal{D}_{n-1}, \bm{\Theta}_{n-1} \sim \mathcal{N}\left(\hat{\mu}_{n-1}(\bm{x}), \hat{\sigma}^2_{n-1}(\bm{x})\right)$ 
    \Else
        \State $\bm{\Theta}_{n-1} \leftarrow \bm{\Theta}_{n-2}$
    \EndIf
    \State \textbf{2. Active Learning acquisition:}
    \State \quad $\bm{x}_{n} \leftarrow \underset{\bm{x}}{\operatorname{arg max}} \: \alpha\left(\bm{x}, \bm{Q}_{n-1},  \hat{\mu}_{n-1}(\bm{x}), \hat{\sigma}_{n-1}(\bm{x})\right)$
    \State \textbf{3. Physical evaluation and augmentation:}
    \State \quad Measure output $Q_{n}$ and noise $\sigma^2_{\varepsilon,n}$ at $\bm{x}_{n}$ 
    \State \quad $\mathcal{D}_{n} \leftarrow \mathcal{D}_{n-1} \cup \{(\bm{x}_{n}, Q_{n}, \sigma^2_{\varepsilon,n})\}$
    \State \quad $f(\bm{x}) \mid \mathcal{D}_{n}, \bm{\Theta}_{n-1} \sim \mathcal{N}\left(\hat{\mu}_{n}(\bm{x}), \hat{\sigma}^2_{n}(\bm{x})\right)$
    \State $n \leftarrow n + 1$
\EndWhile
\State \textbf{4. Optimum estimation:}
\State $\bm{x}^* \leftarrow \underset{\bm{x}}{\operatorname{arg max}} \: \hat{\mu}_{n}(\bm{x})$
\State \Return $\bm{x}^*$ \Comment{Final estimated global optimum}
\end{algorithmic}
\end{algorithm}

Consequently, the \ac{BO} framework can be atomized into three iterative components, followed by a final estimation step. First, inside the loop, the surrogate model (in this case, the heteroskedastic \ac{GP}) and its hyperparameters are updated to approximate the physical system. The second component is the active learning acquisition function, which determines the next optimal control setting by using the surrogate's predictions. Third, the physical oracle is evaluated at the proposed configuration and the training dataset is augmented with the new observation. Once the evaluation budget is exhausted, the iterative loop terminates. Finally, a terminal estimation step evaluates the resulting surrogate to locate the predicted global optimum of the objective function. Refer to Alg.~\ref{alg:hbo} for the formal mathematical framing of this process.
 
Active learning \cite{krause2025probabilistic} governs the sequential decision-making process by optimizing the acquisition function $\alpha(\cdot)$. At iteration $n$ of the optimization loop, the next experimental configuration is selected by maximizing the following proxy function, 
\begin{equation}
    \bm{x}_{n} = \underset{\bm{x}}{\operatorname{arg max}} \: \alpha \left(\bm{x}, \bm{Q}_{n-1}, \hat{\mu}_{n-1}(\bm{x}), \hat{\sigma}_{n-1}(\bm{x}) \right).
\end{equation}
Here, the subscript $n$ indicates that the predictive mean $\hat{\mu}_n(\bm{x})$ and predictive variance $\hat{\sigma}_n^2(\bm{x})$ are evaluated using the \ac{GP} surrogate trained on the observations $\mathcal{D}_n$ available up to the current iteration. Similarly, $\bm{Q}_n$ denotes the observed targets up to such iteration.

We use the Upper Confidence Bound~\cite{shahriari2015taking, ghorbani2024active}, which is a standard acquisition function that favors regions with high predicted values or large uncertainty,
\begin{equation}
    \alpha\left(\cdot \right) = \hat{\mu}_{n-1}(\bm{x}) + \kappa \hat{\sigma}_{n-1}(\bm{x}). \label{eq:acq_function}
\end{equation}
The scalar parameter $\kappa$ is heuristically tuned to control the explicit trade-off bias between exploitation (trusting $\hat{\mu}_n$) and exploration (trusting $\hat{\sigma}_n$). The acquisition function selects the most informative points solely based on the \ac{GP}'s predictive distribution, making noisy spikes in past observations not affect robustness in the exploration-exploitation strategy. Other commonly used acquisition functions in standard \ac{BO} literature, like the Expected Improvement \cite{hase2018phoenics,wu2024learning,ji2022active}, evaluate potential relative to the single best observation found so far. In a heteroskedastic landscape, localized high-noise regions can produce artificially inflated measurements, which misleads the optimization algorithm by dampening the perceived value of unexplored regions.
%\vspace{-0.4cm}
\subsection{Proposed optimization framework}
%\vspace{-0.4cm}
We apply the BO loop to the trapped ion calibration problem by treating the experiment as an unknown mapping from control parameters to score. Figure~\ref{fig:drawing} visually represents this framework alongside Alg.~\ref{alg:hbo}. At each iteration, the calibration sequence is evaluated at a candidate parameter vector, and a finite-shot measurement provides a score. The \ac{GP} surrogate then updates its posterior distribution over the calibration landscape, which is used to find the optimal parameters that maximize the estimation of the score. An acquisition function selects the next set of controls to correct the surrogate and the loop continues until convergence. By doing so, we eliminate the need for a complex physics-based model and calibrate the quantum system using only a limited number of noisy measurements. 

The heteroskedastic noise model provides essential variance information to the \ac{GP}. In this case we assume that measurement noise is dominated by quantum projection noise which, as discussed in more detail in Sec.~\ref{sec:methods:quantum}, depends on the score value and the number of measurement shots $N$. The reader should note that for the present analysis the convergence criterion is based on a heuristically set maximum number of iterations. In this scenario, the optimization step from Fig.~\ref{fig:drawing} is only necessary to track the convergence of the optimum in Sec.~\ref{sec:results}. Please refer to Sec.~\ref{sec:app:param} for a detailed explanation on the \ac{BO} hyperparameter tuning.
%\vspace{-0.4cm}
\section{Quantum System Description}\label{sec:methods:quantum}
%\vspace{-0.4cm}
The optimization procedure described in Sec.~\ref{sec:methods:BO} is independent of the quantum dynamics and classical control infrastructure. Here, we demonstrate its efficacy in calibrating a trapped-ion entangling gate in both simulation and experiment. 
%#######
\subsection{The M\o lmer S\o rensen Gate}
%\vspace{-0.4cm}
Trapped-ion processors encode qubits in long-lived internal electronic states of ions that share collective motional modes~\cite{Leibfried2003}. The most common way to entangle two ions, is through the \ac{MS} interaction~\cite{Sorensen1999,sorensen2000}. Two ions with a qubit transition frequency $\omega_{\mathrm{cl}}$ and a shared motional mode with frequency $\omega_m$ can be entangled with a bichromatic laser field with frequencies $\omega_{b,r} = \omega_{\mathrm{cl}} \pm (\omega_m + \delta)$ and Rabi frequency $\Omega$ applied to each ion. Under the rotating wave and Lamb-Dicke approximations, the \ac{MS} interaction Hamiltonian is
\begin{equation}
H_{MS} = \hbar  \Omega (
\hat{a} e^{-i\delta t} + 
\hat{a}^\dagger e^{i\delta t} )
(\eta_m^{(1)}\hat{\sigma}_{\phi}^{(1)} + \eta_m^{(2)}\hat{\sigma}_{\phi}^{(2)}),
\label{eq:ms_hamiltonian}
\end{equation}
where $\eta_{j,m} = \hat{k}\sqrt{\hbar/2m_\textrm{ion}\omega_m}b_{j,m}$. The laser wave vector, $\hat{k}$, is assumed to always be projected along the desired mode and $b_{j,m}$ is the participation of ion $j$ in mode $m$. Assuming optimal values of $\Omega$ and $\delta$, for a given gate time $t$, the gate will reduce to the ideal unitary,
\begin{equation}
    \hat{U}_{\mathrm{ideal}}
    =
    \exp\left[
    i\frac{\theta}{2}
    \hat{\sigma}_{\phi}^{(1)}
    \hat{\sigma}_{\phi}^{(2)}
    \right].
    \label{eq:ms_ideal_unitary}
\end{equation}
From the simplified Hamiltonian in Eq.~\ref{eq:ms_hamiltonian}, it is possible to analytically calculate the optimal control parameters for a given gate time. However, in practice, many of the assumptions used to arrive at the simplified form do not hold, such as the effect of  pulse shape, off-resonant carrier excitation, or off-resonant excitation of other motional modes. Moreover, the ion qubit and motional frequencies, $\omega_{\mathrm{cl}}$ and $\omega_m$  and the effective Rabi frequency $\Omega$ may drift. Thus, at any given time there is an apriori unknown set of optimal laser powers, frequencies, and phases to implement the desired unitary. 

Here we aim to implement a gate with $\hat{\sigma}_{\phi} = \hat{X}$ and $\theta = \pi/2$ to prepare a maximally entangled state from the ground state, $\ket{gg}$.  Our goal is to find the optimal parameter set
\begin{equation}
    \bm{x}^*
    =
    \left(
    \Omega^*,
    \delta^*,
    \omega_{\mathrm{cl}}^*
    \right),
    \label{eq:parameter_set}
\end{equation}
for a given gate time $t$. We separate $\delta$ and $\omega_{\mathrm{cl}}$ to distinguish error channels~\cite{innsbruckBayesian}. Sideband-detuning errors, $\delta = \delta^* + \Delta{\delta}$, primarily control the residual spin-motion displacement at the end of the gate whereas center-line detuning errors, $\omega_{\mathrm{cl}} = \omega_{\mathrm{cl}}^* + \Delta{\omega_{\mathrm{cl}}}$, track qubit-frequency or laser-frequency drift and generate coherent errors such as unwanted relative phases and residual computational-basis populations~\cite{MartinezGarcia2022}. Errors in $\Omega$, $\Omega/\Omega^*\neq 1$, will yield an over- or under-rotation of the entanglement gate ~\cite{brownFF}. In simulation we assume that $\Omega$ is independent of frequency, although this is a calibration parameter in experiment. In both simulation and experiment we assume $\Omega$ is independent of ions and that $\phi$ is known~\cite{innsbruckBayesian}. 
\begin{figure*}[htp]
    \centering
    \includegraphics[width=\linewidth]{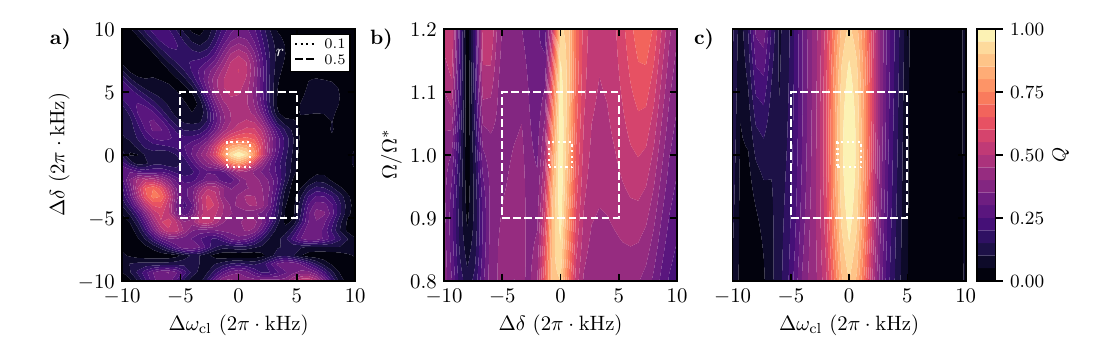}

    \caption{\textbf{Two-parameter calibration landscapes for the two-$\mathrm{MS}_{\pi/2}$ sequence.} Each panel plots the infinite-shot gate score $Q_{\infty}$ while two controls are varied and the remaining controls are fixed at their optimal values. Brighter regions correspond to higher, more optimal scores. The panels depict variations in: \textbf{(a)} sideband detuning and center-line detuning, \textbf{(b)} Rabi frequency and sideband detuning, and \textbf{(c)} Rabi frequency and center-line detuning.}
    \label{fig:contour}
\end{figure*}
%\vspace{-0.4cm}
\subsection{Calibration Sequence and Score Choice}
%\vspace{-0.4cm}
The choice of calibration sequence, measurement, and score will affect the speed and accuracy of any optimization protocol.  Concatenating multiple \ac{MS} gates will increase sensitivity to small control errors but also create a more complicated landscape, including local maxima that do not coincide with the ideal parameter combinations, making convergence to the true optimum unreliable for large initial parameter uncertainties. Thus, there is generally a tradeoff between the local sensitivity of a sequence and if it is globally well-conditioned. 

Each evaluation of the calibration sequence estimates a set of state probabilities $\bm{p}=(P_{gg},P_{ge},P_{eg},P_{ee})$ from a finite number of projective measurements. In the absence of experimental noise, the variance of the probability estimates follows a multinomial distribution~\cite{QPN} with minimal variance when the underlying distribution is at an extreme, e.g. $P_{ij}
\in \{0,1\}$. Thus, it is also important that the calibration sequence ends with both qubits in a computational basis state.

Here, we choose a simple two-gate calibration sequence. We initialize the qubits in $\ket{gg}$, apply two gates with the chosen set of parameters, and measure the state of each ion. Repetition of this sequence $N$ times provides a set of measured probabilities $\bm{p}$. 

For our chosen calibration sequence, all information is held in $P_{ee}$ with unavoidable quantum projection noise~\cite{QPN}. Thus, the score and noise model for the active learning protocol results presented below are
\begin{equation}
Q = P_{ee}, \ \ \sigma^2_{\varepsilon} = \frac{Q(1-Q)}{N}.
\label{eq:Q_noise}
\end{equation}  
% No other noise sources are modeled.

\section{Results}\label{sec:results}
In this section, we validate our optimization approach in simulation and experiment. In simulation, we quantify how accurately the surrogate approximates the landscape as a function of the number of training points and the number of shots per point. We also run the full active learning loop and examine convergence of the calibrated infidelity, first as a function of the initial search-domain scale and then as a function of the finite-shot budget. Finally, we carry out an experimental validation on the protocol, calibrating four control parameters of a two-ion entangling gate.
%#######
\subsection{Simulation Details}
%#######
First, we implement our active learning protocol in simulation, which gives access to a ground-truth $Q_{\infty}$ landscape against which we can benchmark our surrogate model. This is not accessible in an experimental quantum system. 

We use \texttt{IonSim.jl}~\cite{ionSim}, a simulation package that translates hardware parameters into trapped-ion dynamics. We choose to simulate a $100~\mu\mathrm{s}$ square gate on two ${}^{40}\mathrm{Ca}^{+}$ ions using the center-of-mass radial mode with a frequency of 3\,MHz. However, the method is agnostic to the underlying physics. We use the Lamb-Dicke and rotating wave approximations and neglect excitation of other motional modes. We do model off-resonant carrier excitation and the effect of the square pulse excitation. We do not simulate decoherence or dephasing. We apply a small random offset to the parameter bounds for each trial so that the optimum does not always lie at the center of the search domain. Fig.~\ref{fig:contour} shows the simulated $Q_{\infty}$ landscape of our chosen calibration sequence for the parameter region of interest. We see that the score provides high sensitivity to both frequency parameters, $\omega_{\mathrm{cl}}$ and $\delta$, with no secondary maxima of the same height as the optimal. This sequence is fairly insensitive to changes in $\Omega$. It is likely that a more optimal gate sequence could be derived, though the optimality of the sequence would depend on a strong forward model of the system.
%#######
\subsection{Efficacy of the Gaussian Process Surrogate Model} \label{sec:results:GP}
%#######
We begin with a study of the surrogate's ability to globally approximate the quantum landscape as a function of the initial training points,  $m$, and the number of measurements, $N$ at each training point. This analysis provides a preliminary understanding of the surrogate's global accuracy when conditioned on uninformed, randomly sampled data. We compare this static baseline against an active learning approach in  Sec.~\ref{sec:results:opt_1}.
\begin{figure}[ht]
    \centering
    \includegraphics[width=\linewidth]{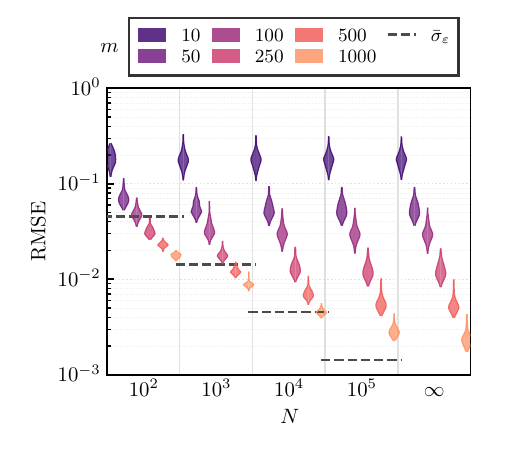}
    \caption{\textbf{Root Mean Square Error distributions of the \ac{GP} predictions relative to the true landscape.} The surrogate's performance is evaluated across different numbers of measurement shots $N$ (grouped on the x-axis) and varying numbers of training points $m$. The errors are displayed as violin plots. The dashed gray lines correspond to the mean binomial noise $\overline{\sigma}_{\varepsilon}$ for each $N$ and provide a reference of the expected uncertainty.}
    \label{fig:GP_fit_RMSE}
\end{figure}
For each $N$, the surrogate is fit to a set of randomly chosen control parameters of size $m$. First, the hyperparameters are optimized via Eq.~\eqref{eq:MLE}, and then the posterior distribution, conditioned on the training data, is evaluated using Eqs.~\eqref{eq:posterior} and \eqref{eq:posterior_mean_s}. To account for statistical variations in the spatial distribution of the training data, this process is repeated for 100 independent random seeds for each set of $(m, N)$. 

We then test each trained model by calculating the \ac{RMSE} over $M=5000$ testing points, disjoint from the training points for a specific seed,
\begin{equation}
    \mathrm{RMSE} = \sqrt{\frac{1}{M}
    \sum_{i=1}^{M}
    \bigl[\hat{\mu}(\bm{x}_i) - Q_{\infty}(\bm{x}_i)\bigr]^2},
\end{equation}
as shown in Fig.~\ref{fig:GP_fit_RMSE}. While the error evaluates the discrepancy between the posterior mean of the \ac{GP} $\hat{\mu}$ and the ground-truth $Q_{\infty}$ of the system, the \ac{GP} is trained on noisy scores at a fixed $N$ value. 
Fig.~\ref{fig:GP_fit_RMSE} shows three main points. First, we compare the \ac{RMSE} to the mean noise model for all test points at a particular $N$, 
\begin{equation}
    \overline{\sigma}_{\varepsilon} = \frac{1}{M}
    \sum_{i=1}^{M}\sigma_{\varepsilon}(\bm{x}_i,N),
\end{equation}
which is indicated with dashed gray lines at each $N$. This expected uncertainty of the quantum system should serve as a baseline for the \ac{GP}'s ability to accurately estimate the true landscape. Interestingly, for large $m$, some \ac{RMSE} distributions lie below $\overline{\sigma}_{\varepsilon}$, indicating that the \ac{GP} mean can denoise the system. Since binomial projection noise is uncorrelated from point to point, the \ac{GP} effectively filters it out by enforcing spatial smoothness across the landscape. 

Secondly, the narrow spread of the violin plots indicates that the global fit is highly robust to the specific locations of the $m$ randomly chosen training points. Third, increasing $N$ does reduce noise and thus increase the accuracy of the model. However, even with abundant data, the global accuracy errors remain bounded in the order of $10^{-2}$ to $10^{-3}$, meaning that $Q$ scores above 0.999 cannot be precisely estimated with random sampling, motivating our active learning strategy. Additional details on the \ac{GP} fit study are provided in Sec.~\ref{sec:app:GP}. 
%\vspace{-0.4cm}
\subsection{Effect of Parameter Uncertainty Range}\label{sec:results:opt_1}
\begin{figure}[t]
    \centering
    \includegraphics[width=\linewidth]{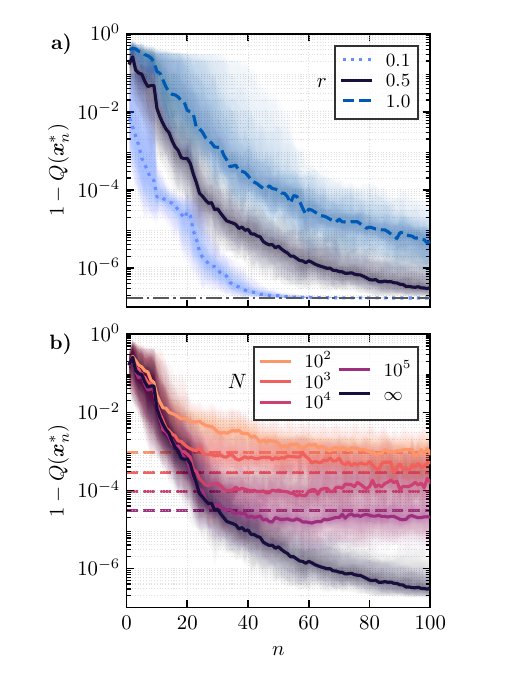}
    \caption{\textbf{Results of \ac{BO} with active learning.}
    \textbf{(a)} Deterministic optimization trajectories for: $r = 1.0$, dashed black, spans $\pm 2\pi\times10$\,kHz in $\omega_{\mathrm{cl}}$ and $\delta$, and $\pm 20\%$ in laser amplitude. The half-scale ($r = 0.5$, solid dark blue) and one-tenth-scale ($r = 0.1$, dotted blue) cases reduce these bounds by factors of two and ten, respectively. The vertical axis shows the optimization loss, $1 - Q_{\infty}$. The bold curves show the median trajectory and shaded trajectories span the 5th to 95th percentile. The gray dashed-dotted line represents the simulation fidelity limit.
    \textbf{(b)} Optimization trajectories with finite-shot sampling evaluated at $r = 0.5$. Trajectories are shown for $N = 10^2$ (light orange), $10^3$ (light red), $10^4$ (pink), and $10^5$ (purple) shots per evaluation. The solid dark blue curve ($N = \infty$) is the deterministic limit. The dashed horizontal lines show shot-dependent convergence limit.}
    \label{fig:result_trends}
\end{figure}

%\vspace{-0.4cm}
Having established the global prediction capabilities of the \ac{GP} for our system, we study how the surrogate benefits from introducing the full active learning loop presented in Fig.~\ref{fig:drawing} and Sec.~\ref{sec:methods:BO} for efficient, low-cost optimization. We first isolate the effect of the initial parameter range of uncertainty by optimizing directly with $Q_{\infty}$. Figure~\ref{fig:result_trends}(a) displays the deterministic score at the optimizer's recommended control vector, $\bm{x}^*_n$, as a function of optimization iteration, $n$, for three search-domain scales. The full-scale domain ($r=1.0$, dashed trace) spans $\pm 2\pi\times10~\mathrm{kHz}$ in the sideband and center-line detunings and $\pm 20\%$ in $\Omega$; the solid ($r=0.5$) and dotted ($r=0.1$) traces reduce these bounds by a factor of two and ten, respectively. For each experimental condition, we run the optimizer 100 times with a different random set of starting points. The lines plot the median optimization path as a function of optimization round $n$, while the shaded regions show the distribution of trajectories spanning the 5th to 95th percentiles.
We see that reducing the uncertainty range both decreases the time to find the optimal parameter set and the variance over initial starting points. This is because a reduced initial parameter range simplifies the landscape and lets the \ac{GP} surrogate identify the high-score region within fewer iterations. At the full scale, secondary high-score regions make the posterior more sensitive to the random initial samples and slow convergence toward the global optimum. 
\begin{table}[!ht]
\centering
\caption{Parameter convergence results at last iteration $n=100$ (median [5th, 95th] percentiles).}
\label{tab:final_param_convergence}
{\textbf{a)} Parameter search scale ($N=\infty$)}
\resizebox{\columnwidth}{!}{%
\begin{tabular}{l c c c}
\hline\hline
$r$ & $\Delta \omega_{\text{cl}}$ ($2\pi \cdot$Hz) & $\Delta \delta$ ($2\pi \cdot$Hz) & $\Omega/\Omega^*$ \\
\hline
$0.1$ & $-0.00$ $[-0.08, 0.06]$ & $-0.29$ $[-0.36, -0.20]$ & $1.00$ $[1.00, 1.00]$ \\
$0.5$ & $0.04$ $[-0.77, 0.76]$ & $-0.30$ $[-2.08, 0.82]$ & $1.00$ $[1.00, 1.01]$ \\
$1.0$ & $-0.38$ $[-9910, 5.51]$ & $0.68$ $[-3412, 8998]$ & $1.01$ $[1.00, 1.20]$ \\
\hline\hline
\end{tabular}}

\vspace{0.5em}

{\textbf{b)} Shot budget ($r=0.5$)}
\begin{tabular}{l c c c}
\hline\hline
$N$ & $\Delta \omega_{\text{cl}}$ ($2\pi \cdot$Hz) & $\Delta \delta$ ($2\pi \cdot$Hz) & $\Omega/\Omega^*$ \\
\hline
$10^2$ & $-4.12$ $[-77.0, 76.0]$ & $12.3$ $[-73.8, 82.1]$ & $1.01$ $[0.97, 1.04]$ \\
$10^3$ & $-1.96$ $[-45.6, 51.7]$ & $-4.71$ $[-55.2, 51.6]$ & $1.00$ $[0.98, 1.03]$ \\
$10^4$ & $0.63$ $[-28.8, 35.4]$ & $2.23$ $[-45.7, 38.6]$ & $1.01$ $[0.98, 1.02]$ \\
$10^5$ & $0.41$ $[-10.2, 7.81]$ & $0.42$ $[-13.4, 12.4]$ & $1.00$ $[1.00, 1.01]$ \\
$\infty$ & $0.04$ $[-0.77, 0.76]$ & $-0.30$ $[-2.08, 0.82]$ & $1.00$ $[1.00, 1.01]$ \\
\hline\hline
\end{tabular}
\end{table}
In the full-domain case, the optimizer robustly finds the global optimum in the median case, with poorly converged seeds appearing only in the extreme percentiles. For the restricted search spaces ($r=0.1$ and $0.5$), the surrogate escapes the local optima regardless of the initial landscape information. Table~\ref{tab:final_param_convergence}(a) gathers the distribution of the optimal control settings at the final optimization iteration for each search-domain scale. Note that the frequency parameters ranges are reported in Hz, showing that the GP finds the optimal frequencies more precisely than $\Omega$, as expected from the topology of the landscape. 
%#######
\subsection{Optimization Convergence with Quantum Projection Noise}\label{sec:results:opt_2}
%#######
Next, we include finite-shot sampling in the score evaluations. The surrogate and the recommended control vector $\bm{x}^*_n$ are built from noisy scores sampled with $N$ shots, but we validate the calibration by evaluating $\bm{x}^*_n$ on the noiseless $Q_\infty$. Fig.~\ref{fig:result_trends}(b) displays $Q_{\infty}$, using $\bm{x}^*_n$ for each $n$, given 100 random initializations, so that the reported infidelity reflects the true gate quality rather than the measurement noise. The initial search-domain is fixed to a moderate bound ($r=0.5$) and a range of finite shot budgets is tested, alongside the ideal infinite-shot limit.

Fig.~\ref{fig:result_trends}(b) shows that the optimizer fidelity saturates at a level set by the number of shots per evaluation, $N$. For $N=100$, the calibrated infidelity plateaus at $1-Q(\bm{x}_{n\to\infty}^*)=1.1\times10^{-3}$, with a 95\% CI of $[0.91, 1.27] \times 10^{-3}$. Larger shot budgets reduce the floor, while the deterministic case continues to improve toward the intrinsic simulation limit. Over the tested range, the saturation level is well described by
\begin{equation}
    1
    -
    Q(\bm{x}_{n\to \infty}^*)
    \approx
    (1
    -
    Q_{\infty}(\bm{x}^*))
    +
    \gamma N^{-1/2},
\end{equation}
where $1-Q_{\infty}(\bm{x}^*)$ is the infidelity floor in the infinite-shot limit, and $\gamma=(1.04\pm0.26)\times10^{-2}$ is a fit parameter and likely depends on the exact calibration sequence and score. This scaling is consistent with the floor being set by the resolution of individual evaluations. 

\begin{figure}[ht!]
    \centering
    \includegraphics[width=\linewidth]{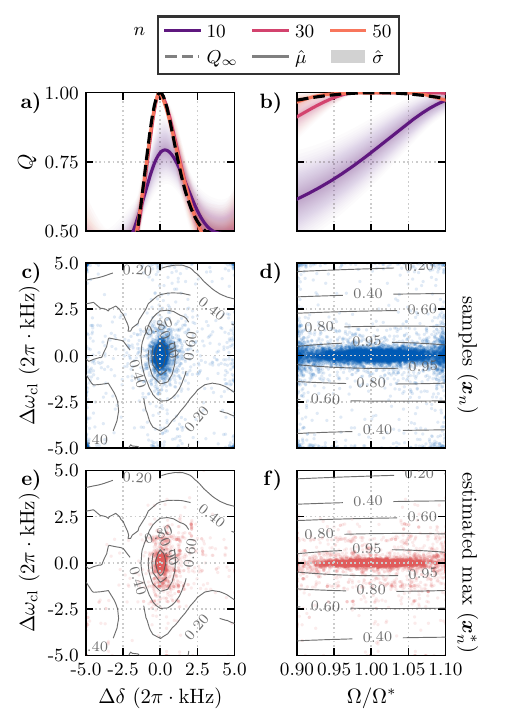}
    \caption{\textbf{Gaussian Process fitting and active learning distributions for the $N=100$ scenario}. (a,b) 1D slices of the predicted score $Q$ along $\Delta\delta$ and $\Omega/\Omega^*$, respectively. The calibration parameters not shown in a given slice are fixed at their optimal values. Solid colored lines and shaded regions correspond to the surrogate's predicted mean $\hat{\mu}$ and uncertainty $\hat{\sigma}$ at different iterations $n$ of the active learning loop. The dashed black line represents the target infinite-shot noiseless landscape $Q_{\infty}$. (c,d) Distribution of candidate points $\boldsymbol{x}_n$ sampled during the active learning step across all iterations (from $1$ to $n_{\mathrm{max}}$) and $100$ independent runs, where denser blue regions indicate parameters sampled at a higher rate. (e,f) The concentration of the corresponding estimated maxima $\boldsymbol{x}^*_n$ for all iterations and runs. For (c-f) gray contour lines provide a reference to the underlying target landscape, corresponding to the 2D slices presented in Fig.~\ref{fig:contour}.}
    \label{fig:gp_results}
\end{figure}

Fig.~\ref{fig:gp_results}(a,b) show examples of one-dimensional cuts through the learned \ac{GP} posterior, when $N=100$, at different points during the active learning process showing the evolution of the model landscape through measurements of the quantum system. The surrogate is able to immediately identify the frequency peaks, and narrows its uncertainty around $\Delta \delta=\Delta \omega_{\mathrm{cl}}=0$. In contrast, the amplitude cut remains broad because the chosen calibration sequence and score changes weakly with respect to perturbations near the optimum, necessitating a higher $N$ to accurately discern small differences. The landscape plots in Fig.~\ref{fig:gp_results}(c,d) show where the active learning algorithm samples across all iterations and independent runs, for $N=100$.  \ac{BO} makes informed decisions about the samples that would increase confidence in the optimal parameters. The landscape plots in Fig.~\ref{fig:gp_results}(e,f) show the evolution of the expected max, $\bm{x}_n^*$, provided by the \ac{GP} surrogate. Given a reasonable experimental parameter range ($r=0.5$), $\bm{x}_n^*$ always converged to a high-fidelity control vector.

Table~\ref{tab:final_param_convergence}(b) summarizes the distribution of the final recommended control parameters for different shot budgets. The optimal amplitude ratio consistently converges to the ideal value of 1.00 with minimal variance across all tested shot budgets and the frequency parameters converge within tens of Hz to the ideal parameter set, even when starting with a kHz level parameter uncertainty. These results demonstrate that a physics-agnostic active learning approach has the ability to readily calibrate a quantum processor, even when optimizing a noisy landscape and with an unknown and non-ideal score landscape.
%#######
\subsection{Experimental Validation}\label{sec:results:exp}
%#######

Our simulations assumed a perfect, if uncalibrated, quantum system. However, any hardware implementation will suffer from decoherence, fast control noise, and state preparation and measurement errors. Here, we test the efficacy of a model that only has knowledge of quantum projection noise by calibrating an entangling gate between two \(^{171}{\rm Yb}^+\) ions using QSCOUT~\cite{QSCOUT}. The underlying gate dynamics are controlled by the same general physics as in the simulation, though the exact implementation differs, as it uses Raman excitation and a balanced Gaussian pulse~\cite{balancedGate}. The calibrated control vector is
\[
\bm{x}^* =
\left(
I^*,
\delta^*,
\omega_{\mathrm{cl}}^*,
\zeta^*
\right),
\]
where $\delta$, and $\omega_{\text{cl}}$ correspond to the frequency parameters calibrated in simulation, $\zeta$ is the ratio of the laser intensities for the red- and blue-sideband lasers. Unlike in simulation, we do not have direct access to $\Omega$, and instead manipulate the logarithmic power scale $I$ applied to an acoustic optic modulator which controls $\Omega$ at the two ions. As above, we use two MS gates as the calibration sequence and measure the population of the $\ket{ee}$ state, denoted as $Q(\bm{x}^*_n)$. The number of shots is fixed to \(N=100\) based on prior estimates of the QSCOUT testbed fidelity~\cite{balancedGate}. The parameter bounds are centered around the manual calibration result. However, no prior knowledge is provided to the model. As in the simulation case, we test two parameter ranges, denoted as $r_\text{exp}$. The full-scale domain $r_\text{exp}=1$ spans $\delta=\delta_0 \pm 5\times2\pi$\,kHz, $\omega_{\text{cl}}=\omega_{\text{cl},0} \pm 5\times2\pi$\,kHz, and $I = I_0\pm10$. In the reduced range, $r_\text{exp}=0.5$, reduces these bounds by a factor of two. The smaller bound, $r_\text{exp}=0.5$,  was motivated based on prior calibration work~\cite{innsbruckBayesian} while the larger parameter range emphasizes the functionality of the method even with a complex extended landscape.

\begin{figure}[ht!]
    \centering
    \includegraphics[width=\linewidth]{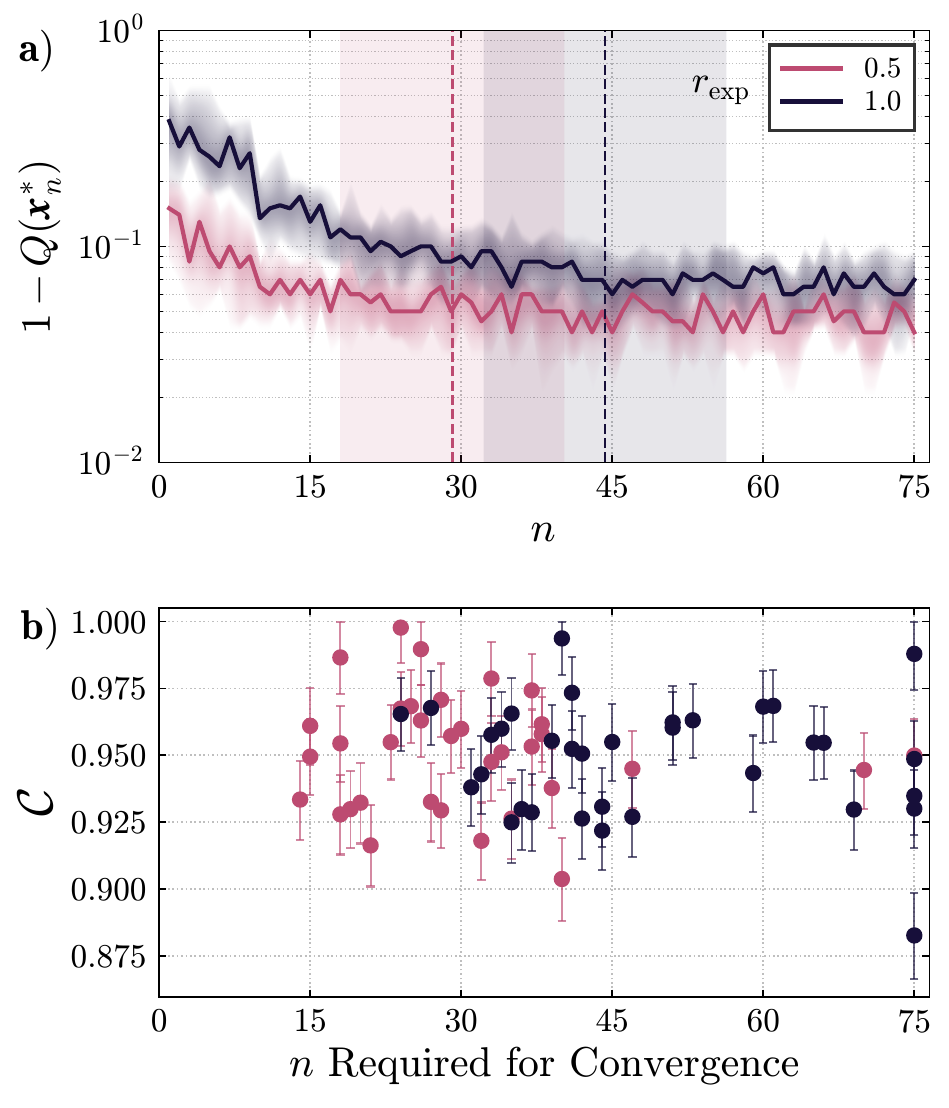}
    \caption{\textbf{Calibration of the MS gate on two ions using model-free Bayesian techniques}. (a) Measured population error $1 - Q(\bm x^*_n)$ after two MS gates, given \(N = 100\) for the standard (pink) and extended (purple) parameter bounds, after each active learning step. The shaded trajectories span the 5th to 95th percentile, and the bold curves show the median trajectories. The dashed vertical lines and corresponding shaded columns represent the average iteration required for convergence and its one-standard-deviation interval for each parameter range. (b) Final parity contrast, $\mathcal{C}$, for each individual run as a function of the active learning iterations required for convergence. Contrast is measured from the parity curves using the recommended parameters after a single MS gate. Runs that do not converge appear at step 75. Error bars represent one standard deviation assuming a binomial distribution.}
    \label{fig:experiment}
\end{figure}

We apply the heteroskedastic Bayesian optimization process validated in simulation, this time on an experimental trapped ion system. For each parameter range we repeat the process for 35 independent random seeds over the course of many hours, without any manual intervention or recalibration.  Before active learning, 15 points are randomly sampled to start building the surrogate model. Fig.~\ref{fig:experiment}(a) shows the median population score at the optimizer's recommended calibration configuration $Q(\bm{x}^*_n)$ across iterations, for the two sets of parameter bounds. The $n=1$ point shows that the 15 points already provide enough information about the landscape for a strong initial guess, with a median $Q(\bm{x}^*_0)$ of 0.85 (0.65) with a 95\% CI of [0.81, 0.89] ([0.51, 0.74]).
The median values after 75 active learning steps, were 0.96 (0.93), with a 95\% CI of [0.94,0.97] ([0.91,0.94]).
The deviation is in line with binomial shot noise as described in Eq.~\ref{eq:Q_noise}. Notably, the spread of $Q(\bm{x}^*_n)$ normalized to the median are comparable in both experiment and simulation. This suggests that the low $Q(\bm{x}^*_n)$ with respect to simulation could be attributed to a missing calibration parameter or unknown system instability, and not to the optimizer itself.

Post-calibration, we independently bound the single-\ac{MS} gate fidelity to validate the calibration protocol. To do this, a single MS gate is run with the final control parameters found by the optimizer. This is followed by a \(\mathcal{R}(\pi/2)\) rotation pulse with varying analysis phase, \(\phi\) where the parity contrast \(\mathcal{C}\) provides a lower-bound on the entanglement fidelity~\cite{Sackett2000Mar}. The full parity measurements used to estimate $\mathcal{C}$ in Fig.~\ref{fig:experiment}(b) are shown in Appendix~\ref{sec:app:Parity}.

Figure~\ref{fig:experiment}(b) shows the effective calibration iterations and the final measured parity. Using standard parameter bounds, the mean $\mathcal{C}$ is 0.952 $\pm$ 0.022 and with the extended parameter bounds, $\mathcal{C}$ is 0.947 $\pm$ 0.025. Although we let the calibration run for 75 steps in the experiment, we also calculate the expected time-to-convergence in a real-time calibration in post processing. We define convergence as the point at which the measured and \ac{GP}-predicted $Q(\bm x^*_n)$ plateau, and the step-to-step fluctuations satisfy the condition $|\Delta Q| < \sigma_\varepsilon$ for 5 consecutive steps, where $\Delta Q = Q_n - Q_{n-1}$ and $\sigma_\varepsilon$ is the theoretical binomial shot noise defined in Eq.~\ref{eq:Q_noise}. Convergence occurs in Fig.~\ref{fig:experiment}(b) after an average of $n=44 \pm 11$ and $n=60 \pm 12$ steps for $r_{\rm exp}=$ 0.5 and 1.0, respectively, when including the 15 initial samples (not shown in Fig.~\ref{fig:experiment}). Prior physics-based algorithms required 1200$\pm$500 shots~\cite{innsbruckBayesian} to complete calibration, when using parameter ranges comparable to $r_{\rm exp} = 0.5$. This means that the cost associated with generalizing the model, is roughly 4$\times$, emphasizing the efficiency of calibrating quantum systems using active learning with a physics-agnostic surrogate model.

 \section{Conclusion}\label{sec:conclusions}
%\vspace{-0.4cm}
In this manuscript, we demonstrate a physics-agnostic active learning protocol for calibrating trapped-ion entangling gates. A Gaussian process surrogate model, which accounts for heteroskedastic measurement noise, paired with Bayesian optimization, learns a noisy calibration landscape and identifies high-fidelity control parameters~\cite{williams2006gaussian,shahriari2015taking}. In simulation, the surrogate captures the three-parameter landscape, the calibrated infidelity converges to a floor set by finite-shot quantum projection noise, and the full protocol shows robustness across a large range of initial control settings. In an experiment on the QSCOUT platform, the protocol successfully calibrates a two-ion gate over four control parameters in $44\pm11$ evaluations of 100 shots each, reaching a parity contrast of $0.952\pm0.022$. The remaining experimental infidelity is consistent with an unmodeled control parameter or fast hardware instabilities.

These results show the trade-off between generality and cost. Physics-based models precompute the probability landscape instead of learning it, which reduces the number of shots to convergence~\cite{innsbruckBayesian} by a  factor of roughly four. Future extensions could encode a lightweight physical expansion to the noise model to help calibration in the presence of experimental error. This could also help when scaling to more parameters or more qubits. Multi-fidelity approaches combining classical simulations, low-shot quantum measurements to approximate the landscape, and high-fidelity measurements near the optimum could reduce the total shot budget while maintaining robustness. Finally, the learned model could be reused and re-optimized for faster drift maintenance. Together, these directions validate surrogate-based active learning as a routine component of calibration, control, and verification as quantum systems scale.

\section{Data and Code Availability}

The data and code supporting the findings of this study are openly available at \url{https://github.com/PatGCaspuenas/QuantumFidelityOptimization}.

\section{Acknowledgements}
This work was funded in part through NSF award OMA-2329020 and through the NSF AI Institute in Dynamic Systems grant number 2112085. C.W. was supported in part through the Advancing Quantum-Enabled Technologies (AQET) traineeship program at the University of Washington through NSF award DGE-2021540. FZ acknowledges the support of the JRC STEAM STM-Politecnico di Milano agreement. The experiments were run on the QSCOUT testbed, funded in part by the U.S. Department of Energy, Office of Science, Office of Advanced Scientific Computing Research Quantum Testbed Program. Sandia National Laboratories is a multimission laboratory managed and operated by National Technology \& Engineering Solutions of Sandia, LLC, a wholly owned subsidiary of Honeywell International Inc., for the U.S. Department of Energy’s National Nuclear Security Administration under contract DE-NA0003525. SAND2026-28555O.  This paper describes objective technical results and analysis. Any subjective views or opinions that might be expressed in the paper do not necessarily represent the views of the U.S. Department of Energy or the United States Government.

\bibliography{main} \newpage
\section{Appendix}\label{sec:appendix}
%#######
\subsection{Bayesian Optimization hyperparameter choices}\label{sec:app:param}
%#######
This section justifies the selection of several empirically tuned parameters within the \ac{BO} pipeline. Initial training dataset sizes of $m=12$ for Sec.~\ref{sec:results:opt_1} and \ref{sec:results:opt_2}, and $m=15$ for Sec.~\ref{sec:results:exp}, were selected following the empirical rule of \cite{booker2007movars} where $m\approx\frac{(d+1)(d+2)}{2}$, being $d$ the dimensionality of the input vector. Increasing the initial training set within that order of magnitude did not show significant changes in the convergence of the fidelity error across any of the evaluated noise levels.

For the acquisition function defined in Eq.~\ref{eq:acq_function}, the parameter $\kappa$ was set to 1.96 to focus on the $95\%$ confidence interval of the \ac{GP} predictions. This choice established a balanced exploration-exploitation search strategy.

Finally, while new training points were added to the \ac{GP} posterior at every iteration, the hyperparameters were only updated every $h=10$ iterations (see Alg.~\ref{alg:hbo}). Freezing the kernel length scales for this number of iterations and updating them periodically led to more robust and accurate results than updating them at every iteration. Given the small number of points added to the \ac{GP} in a 3D space (up to 100 points), freezing the hyperparameters for an extended number of iterations was considered detrimental to capturing the information of the landscape. To prevent ill-conditioning of the covariance matrix, a minimum distance between the point to be added and the existing ones was set at $10^{-4}$.
%#######
\subsection{Additional metrics for Gaussian Process fit study}\label{sec:app:GP}
\begin{figure}[h]
    \centering
    \includegraphics[width= 0.8\linewidth]{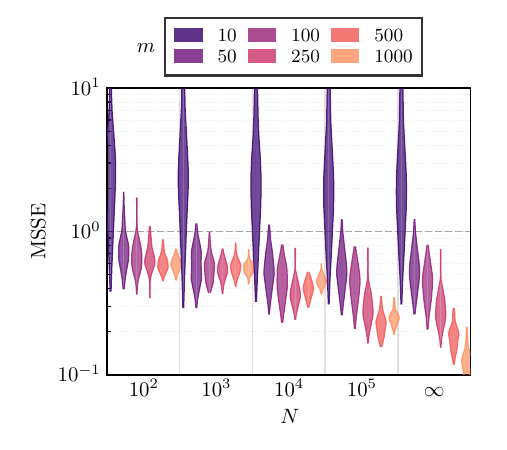}
    \caption{\textbf{Mean Squared Scaled Error distributions of the Gaussian Process predictions relative to the target infinite-shot noiseless landscape $Q_{\infty}$}. The surrogate's performance is evaluated across different numbers of measurement shots $N$ (grouped on the x-axis) and varying numbers of training points $m$ (color-coded), which are randomly distributed across the 3D input space. The errors are displayed as violin plots. The dashed gray line at MSSE $= 1$ serves as a reference for ideal uncertainty quantification.}
    \label{fig:GP_fit_MSSE}
\end{figure}

%#######
Following the same surrogate fitting process as in Sec.~\ref{sec:results:GP} for each $(m, N)$ configuration, we can also analyze the \ac{MSSE} distribution, where the metric is given by
\begin{equation}
    \mathrm{MSSE} = \frac{1}{M}
    \sum_{i=1}^{M}
    \frac{\bigl[\hat{\mu}(\bm{x}_i) -
          Q_{\infty}(\bm{x}_i)\bigr]^2}
         {\hat{\sigma}^2(\bm{x}_i)}. \label{eq:MSSE}
\end{equation}
In Eq.~\ref{eq:MSSE}, the reliability of the \ac{GP}'s uncertainty $\hat{\sigma}$ of the noise-free landscape is evaluated. The actual error between the \ac{GP} posterior mean and the target infinite-shot $Q_{\infty}$ is expected to be within the \ac{GP}'s variance bounds, making $MSSE=1$ the optimal metric. Figure~\ref{fig:GP_fit_MSSE} depicts the \ac{MSSE} distribution across different values of $m$ and $N$.
In Fig.~\ref{fig:GP_fit_MSSE}, we can observe two main trends. First, there is a clear jump from $m=10$ to $m=50$, where the \ac{GP} transitions from being overconfident to underconfident in its predictions. For extreme data sparsity ($m=10$), the uncertainty of the surrogate does not cover the actual error of the posterior mean with respect to the noiseless landscape, regardless of the noise level. The second observation is the non-monotonic trend of the \ac{MSSE} as $m$ increases under finite-shot noise. In the infinite-shot limit, the \ac{GP} improves its posterior mean predictions faster than it narrows its uncertainty. This shows that the model is always conservative in the regions where data is not available, making it more uncertain about its predictions than it should be globally. At finite $N$, the noise of the system introduces an additional fitting error that becomes more evident as the surrogate is conditioned on more data. This error source compensates the posterior variance of the \ac{GP} and slightly pushes \ac{MSSE} towards 1.
\begin{figure*}[htb]
    \centering
    \includegraphics[width=0.8\linewidth]{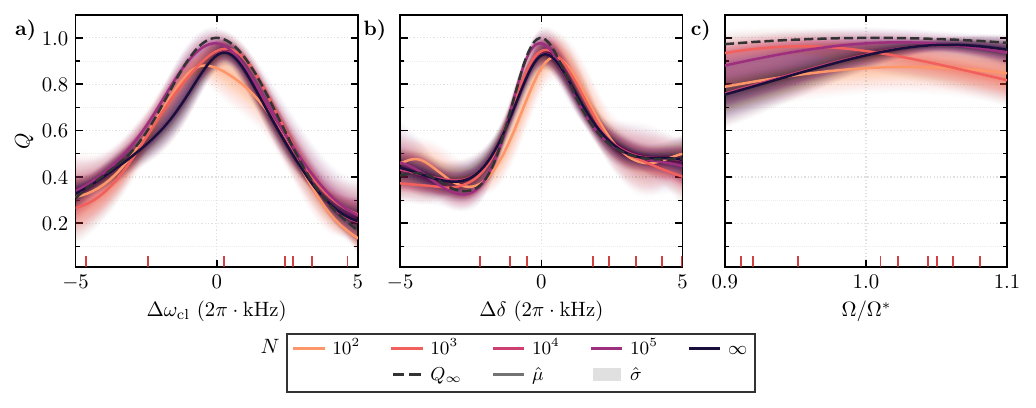}
    \caption{\textbf{1D cross-sections of Gaussian Process mean and variance predictions using $m=50$ training points across a 3D input space, evaluated for different numbers of measurement shots $N$}. Panels (a), (b), and (c) display slices of the predicted score $Q$ along the center-line detuning $\Delta\omega_{\mathrm{cl}}$, sideband detuning $\Delta\delta$, and normalized Rabi frequency $\Omega/\Omega^*$, respectively. The calibration parameters not shown in a given slice are fixed at $\Delta\omega_{\mathrm{cl}}=0$, $\Delta\delta=0$, or $\Omega/\Omega^*=1$. Solid colored lines and shaded regions correspond to the surrogate's predicted mean $\hat{\mu}$ and uncertainty $\hat{\sigma}$ for each $N$. The dashed black line represents the target infinite-shot noiseless landscape $Q_{\infty}$ to be fitted. Red ticks along the bottom axes indicate the locations of the training points that fall within a $10\%$ distance of the respective 1D slice with respect to the full-domain size.}
    \label{fig:GP_fit_slices}
\end{figure*} 
Figure~\ref{fig:GP_fit_slices} visualizes how the \ac{GP} predictions evolve for a fixed number of training points $m=50$ under varying finite-shot measurement budgets. Despite all $N$ scenarios being conditioned on the same calibration settings, the actual shape of the posterior considerably fluctuates. These variations are more noticeable near the local peaks, which require higher resolution to accurately resolve, and in regions where $Q\approx 0.5$, where the underlying binomial projection noise is maximized. However, even with sparse information, if one of the points lie near the high-fidelity regions, the surrogate retrieves a close optimal location for the calibration parameter prior to any active learning. Optimization of the calibration protocol, so that all target parameters fit this profile, will likely yield extremely efficient and accurate calibration results in this scenario.
%#######
\subsection{Population and Parity Data}\label{sec:app:Parity}
%#######

\begin{figure}[t]
    \centering
    \includegraphics[width=0.8\linewidth]{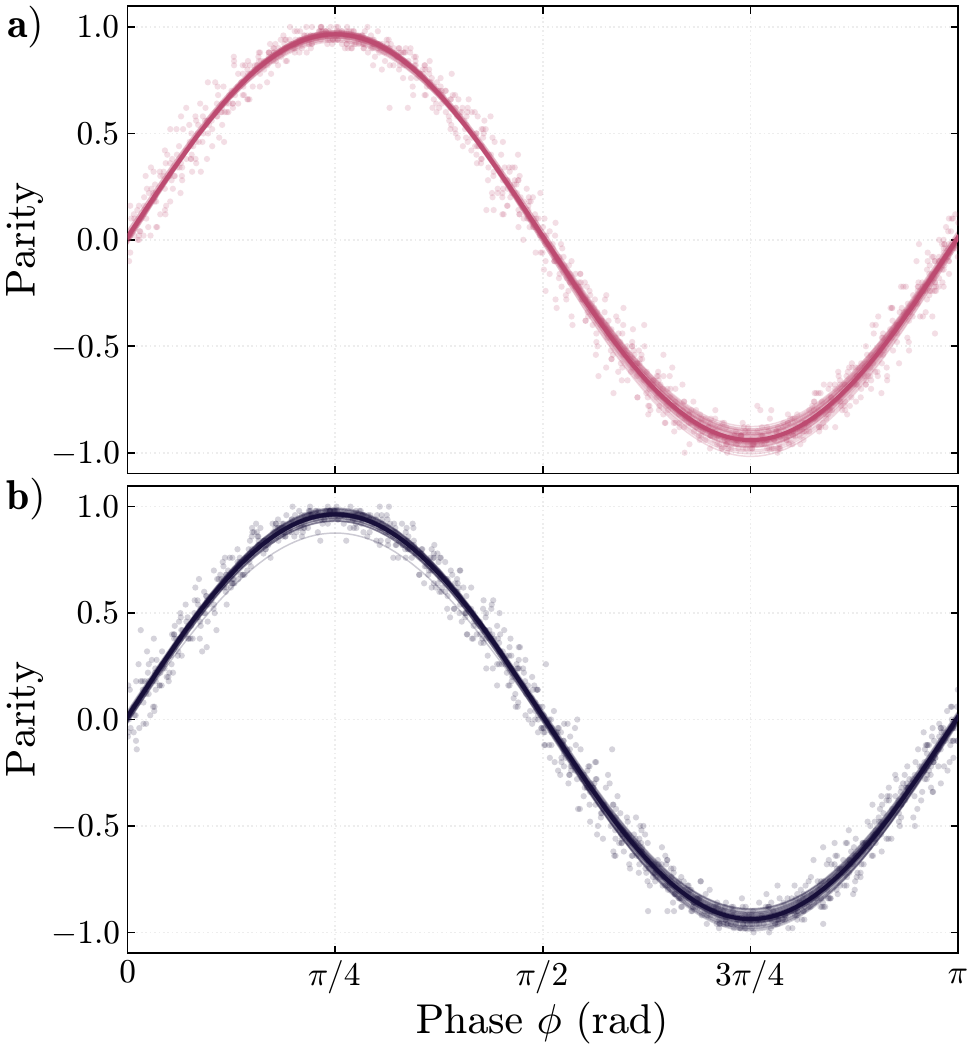}
    \caption{\textbf{Parity measurements taken at the recommended control vector}. Each measurement was taken by applying a single MS entangling pulse, followed by a $\mathcal{R}(\frac{\pi}{2})$ rotation pulse at 10 different control phases between 0 and $\pi$, with $N = 100$ for each measurement. (a) Parity measurements after calibration at the standard range, $r_\textrm{exp} = 1$. (b) Parity measurements after calibration at the extended parameter range, $r_\textrm{exp} = 1$.}
    \label{fig:parity}
\end{figure}

\begin{comment}
Raw measurement values for 35 experimental trajectories are visualized in Fig.~\ref{fig:population_and_model}. The measured $Q(\bm{x}^*_n)$ corresponding to $P_{ee}$ is shown in Fig.~\ref{fig:population_and_model}(a) for both experimental parameter ranges. The $Q(\bm{x}^*_n)$ results predicted by the \ac{GP} are shown in Fig.~\ref{fig:population_and_model}(b). Once the model becomes limited by shot noise, oscillations around the mean of the ensemble, visualized by the bold black line, are observed. It is not expected that with additional active learning steps those results would improve, so, qualitatively, we therefore consider the calibration to have reached convergence once this behavior is seen. This is validated by checking for the point at which this behavior occurs for 5 consecutive steps.
\begin{figure*}[ht!]
    \centering
    \includegraphics[width=\linewidth]{Figures/Appendix/population.pdf}
    \caption{Measurement data for the standard (Left) and extended (Right) calibration ranges over 35 trials. (a) Measured $Q(\bm{x}^*_n)$ for each active learning step. (b) Predicted $Q(\bm{x}^*_n)$ for each active learning step.}
    \label{fig:population_and_model}
\end{figure*} 
\end{comment}
After 75 active learning steps a parity measurement was taken using the recommended control vector. Parity measurements were taken by applying a single MS entangling pulse, followed by a $\mathcal{R}(\frac{\pi}{2})$ rotation pulse at 10 different control phases between 0 and $\pi$, with N = 100 for each measurement. The results for the standard parameter range are shown in Fig. \ref{fig:parity}(a) and for the extended range in Fig. \ref{fig:parity}(b).

\end{document}